\newcommand{\qgas}{Q_g}
\newcommand{\qstar}{Q_s}
\newcommand{\qrw}{Q_{\rm RW}}

\documentclass[a4paper,fleqn,usenatbib]{mnras}

\usepackage{newtxtext,newtxmath}

\usepackage[T1]{fontenc}
\usepackage{ae,aecompl}

\usepackage{graphicx}	
\usepackage{amsmath}	
\usepackage{amssymb}	

\title[Origin of low surface brightness galaxies]
{Origin of low surface brightness galaxies: A dynamical study}

\author[P. Garg and A. Banerjee]
       {Prerak Garg$^{1}$ and
        Arunima Banerjee$^{2}$\thanks{E-mail : arunima@iisertirupati.ac.in} \thanks{This research was carried out at \emph{The Inter-University Centre for Astronomy and Astrophysics, Pune 411007, India} }\\
$^1$  Department of Physics, Savitribai Phule Pune University, Pune 411007, India \\
$^2$  Indian Institute of Science Education and Research, Tirupati 517507, India \\
}

\begin{document}
\label{firstpage}
\pagerange{\pageref{firstpage}--\pageref{lastpage}}
\maketitle
\begin{abstract}
Low Surface Brighness Galaxies (LSBs), inspite of being gas rich, have low star formation rates and are, therefore, low surface brightness in nature.
We calculate Q$_{\rm{RW}}$, the 2-component disc stability parameter as proposed by Romeo \& Wiegert (2011), as a function of galactocentric radius $R$ 
for a sample of five LSBs, for which mass models, as obtained from HI 21cm radio-sythesis observations and R-band photometry, were available in the 
literature. 
We find that the median value of Q$_{\rm{RW}}^{\rm{min}}$, the minimum of Q$_{\rm{RW}}$ over $R$, 
lies between 2.6 and 3.1 for our sample LSBs, which is higher than the median value of 1.8 $\pm$ 0.3 for Q$_{\rm{RW}}^{\rm{min}}$ for a sample
 of high surface brightness galaxies (HSBs) as obtained in earlier studies. This clearly shows that LSBs have more stable discs than HSBs, 
which could explain their low star formation rates and, possibly, their low surface brightness nature. Interestingly, the calculated values of Q$_{\rm{RW}}$ 
decrease only slightly (median Q$_{\rm{RW}}^{\rm{min}}$ $\sim$ 2.3 - 3) if the discs were taken to respond to the gravitational potential
 of the dark matter halo only, but reduce by $\sim$ a factor of 2-3 (median Q$_{\rm{RW}}^{\rm{min}}$ $\sim$ 0.7 - 1.5) if they respond 
to their self-gravity alone. This implies that the dark matter halo is crucial in regulating disc stability in LSBs, which may have important 
implications for models of galaxy formation and evolution.

\end{abstract}
\section{Introduction} 

\noindent Low Surface Brightness galaxies (LSBs) are galaxies with central B-band surface brightness ${{\mu}_B}(0)$
$>$ 22-23 mag arcsec$^{-2}$ (Impey \& Bothun 1997; Schombert et al. 2001);
the faintest LSBs observed so far have ${{\mu}_B}(0)$ $\sim$ 26 mag arcsec$^{-2}$.
LSBs, therefore, are at least two orders of magnitude fainter than the galaxies conforming to Freeman's Law (Freeman 1970), which states that
 most of the disc galaxies have a central B-band surface brightness ${{\mu}_B}(0)$ $\sim$ 21.65 $\pm$ 0.30 mag arcsec$^{-2}$.  
They are ubiquitous, and therefore common, and may be located either in field or in cluster
 environment. Morphologically, LSBs are mostly disc galaxies (existence of a few elliptical LSBs have however been reported in the literature), 
spanning a wide range of luminosities from dwarf-irregulars to giant spirals, and thus encompass a group of galaxies with diverse physical properties 
(de Blok, McGaugh \& van der Hulst 1996).

LSBs are not just faded remnants of high surface brightness galaxies (HSBs) that have ceased to form stars; this was confirmed by the fact that 
no correlation was found to exist between the values of the central surface brightness and colors of LSBs (McGaugh \& Bothun 1994). 
LSBs are, in fact, quite blue in color, often few tenths of a magnitude bluer than HSBs, 
which indicates that they are forming stars even at the current epoch (van der Hulst et al. 1993). 
Besides, they exhibit a low star formation rate (SFR) $\sim$ 0.001 - 0.1 M$_{\odot}$ yr$^{-1}$, which is almost a few orders of 
magnitude lower compared to HSBs (Wyder et al. 2009; Westfall et al. 2014); blue color together with a low star formation rate was
 interpreted as being indicative of a late formation with a slow evolution rate for these galaxies (McGaugh \& Bothun 1994). 
This may appear to be puzzling as LSBs are rich in gas, the fuel for star formation, as indicated by high values of $M_{\rm{HI}}/L_B$ ($\sim$ 1) and 
$M_{\rm{HI}}/M_{\rm{dyn}}$ ($\sim$ 0.1), where $M_{\rm{HI}}$ is the 
total mass in neutral hydrogen gas ($\rm{HI}$), $L_B$ the total blue luminosity, and $M_{\rm{dyn}}$ the dynamical mass of the galaxy 
(de Blok, McGaugh \& van der Hulst 1996).
However, it was argued that although LSBs have rich reserves of $\rm{HI}$ 
by mass, the observed ${\rm{HI}}$ surface density falls below the critical value of the gas surface density required for star formation as given by the 
Kennicutt (1989), an empirical law relating star formation with gas surface density for a large sample of spiral galaxies 
(van der Hulst et al. 1993, Wyder et al. 2009). Interestingly, in a similar study, Hunter et al. (1998) found that the critical gas surface density for star 
formation in dwarf irregulars to be lower by a factor of 2 than that in HSBs. This already implies that the value of the critical gas surface density  for star formation is not universal, and may itself be a function of other physical parameters which could vary with galaxy mass and morphology. 
This is further evident from the existence of other empirical laws determining star formation in galaxies, which have explicit dependence, for example, on the dynamical 
time, in addition to the gas surface density (Boissier \& Prantzos 1999). 

The value of critical gas density in Kennicutt (1989) translates to a Q-value of 
1.4, Q being the Toomre stability criterion (Toomre 1964) indicating the stability of a self-gravitating disc of stars against local, axi-symmetric 
perturbations; Q > 1 denotes a dynamically stable disc and vice-vera. Galactic discs, however, are not single-component systems, especially late-type, gas-rich galaxies, where the gravitational potential due to stars and gas dominate the disc dynamics on an equal footing. 
Besides, being a low dispersion component, gas is constrained to a thin layer near the galactic midplane and therefore may strongly dominate the disc dynamics there, inspite of constituting just a sub-dominant mass fraction of the galaxy (Banerjee \& Jog 2007). 
2-fluid and 2-component stability parameters for a gravitationally-coupled star-gas system in the force field of the dark matter halo has been 
obtained analytically by several authors in the literature (Wang \& Silk 1994, Jog 1996, Romeo \& Wiegert 2011). 
In this paper, we study the disc stability of a sample of LSBs by calculating the 2-component stability parameter, 
Q$_{\rm{RW}}$, as proposed by Romeo \& Wiegert (2011), and compare with the Q$_{\rm{RW}}$ values obtained for a combined sample of HSBs and a few LSBs,  
available in the literature. We also study the dependence of the observed SFR densities for our sample LSBs, as was determined earlier from 
direct observations, on our calculated 
Q$_{\rm{RW}}$ values, and compare with the trend observed in normal spirals, as seen in numerical simulations and observational studies. Finally, we study the relative roles of the 
gravitational potentials of the disc and the dark matter halo in regulating the disc stability in our sample LSBs.

The rest of the paper is organized as follows: in \S 2 we discuss the stability criterion against local axi-symmetric perturbations for a 2-component disc, in \S 3 the sample, in \S 4 the input parameters, in \S 5 results and discussion followed by conclusions in 
\S 6.

\section{Stability criterion for a 2-component galactic disc}

The theory of gravitational instabilities in a galactic disc for                        
2-component galactic disc models of stars and \mbox{H\,{\sc i}} supported by rotation and random motion 
has been studied in detail and several 2-component stability parameters proposed in the past (See \S 1).
In order to compare our results with earlier studies in the literature, in this paper
 we will adopt the stability parameter given by Romeo \& Wiegert (2011), $Q_{\rm{RW}}$, which is given by

\begin{equation}
\qrw^{-1} = \left\{
\begin{array}{ll}
w_{\sigma}/\mathcal{T}_s \qstar + 1/\mathcal{T}_g \qgas & {\rm for}\
\mathcal{T}_s \qstar > \mathcal{T}_g \qgas \\[6pt]
1/\mathcal{T}_s \qstar + w_{\sigma}/\mathcal{T}_g \qgas & {\rm for}\
\mathcal{T}_s \qstar < \mathcal{T}_g \qgas

\end{array}
\right.,
\label{eq:qrw}
\end{equation}

where the 1-component stability parameter is given by
\begin{equation}
Q_i = {\kappa} {\sigma}_{R,i}/(\pi G {\mu}_i ) 
\end{equation}

Here,  ${\kappa}$ is the epicyclic frequency, ${\sigma}_{R,i}$ the radial velocity dispersion and 
${\mu}_i$ the surface density of the $i^{th}$ component where $i = s$ or $g$ corresponding to stars and gas respectively. In other words, the 1-component stability parameter is determined by a balance between the pressure gradient and the centrifugal force from the coriolis spin-up of the perturbation on one hand, and its self-gravity on the other. Besides, $Q_i$, ${\kappa}$, ${\sigma}_{R,i}$ and ${\mu}_i$ are all functions of $R$. The epicyclic frequency ${\kappa}$ at any galactocentric radius $R$ is given by
\begin{equation}
{{\kappa}}^2 = {( R \frac{ {\partial} {{\Omega}_{\rm{Total}}}^2}{\partial R} + 4 {{\Omega}_{\rm{Total}}}^{2} )} 
\end{equation}
where  ${\Omega}_{\rm{Total}}$ is the total angular frequency given by
${\Omega}_{\rm{Total}}^2$ = $\frac{1}{R} \frac{ {\partial} { {\Phi}_{\rm{Total} } } }{\partial  R}$ = ${V_{\rm{Rot}}}^2/{R^2}$. Here
 ${\Phi}_{\rm{Total}}$ is the total gravitational potential, and $V_{\rm{Rot}}$ the rotational velocity at $R$, which is given by the observed rotation 
curve of the galaxy.  Therefore, at any radius $R$, the square of the total angular frequency is obtained by adding in quadrature ${\Omega}_{\rm{Disk}}$, the angular frequency due to the gravitational potential of the disc only, and ${\Omega}_{\rm{DM}}$, that due to the dark matter only i.e., 
\begin{equation}
{\Omega}_{\rm{Total}}^2 = {\Omega}_{\rm{Disk}}^2 + {\Omega}_{\rm{DM}}^2
\end{equation}
We note here that the role of the dark matter halo in regulating the local disc stability lies only 
in its contribution to the net galactic potential, which in turn determines the angular and epicyclic frequencies. \\ 

\noindent $w_{\sigma}$ is the weight of the more stable component and is given by
$$ w_{\sigma} = \frac{{\sigma}_{R,s} {\sigma}_{R,g} }{{\sigma}_{R,s}^2 + {\sigma}_{R,g}^2  } $$

\noindent and $\mathcal{T}_i$, the correction for disc thickness for the i$^{th}$ component, as given by Romeo \& Falstad (2013), is 

\begin{equation}
\mathcal{T}_i = \left\{
\begin{array}{rl}
0.8 + 0.7\ \alpha_i & {\rm for}\ \ 0.5 < \alpha_i \leq 1 \\[6pt]
1.0 + 0.6\ \alpha_i^2 & {\rm for}\ \ 0.0 < \alpha_i \leq 0.5
\end{array}
\right.
\label{eq:thickq}
\end{equation}

such that $\mathcal{T}_i\geq1$. Here $\alpha_i =
\sigma_{z,i}/\sigma_{R,i}$ where ${\sigma}_{z,i}$ is the vertical velocity dispersion of the i$^{th}$ component.
For gas, we assume the velocity ellipsoid is isotropic such that
$\mathcal{T}_g = 1.5$. For stars, we take $\alpha_s =\sigma_{z,s}/\sigma_{R,s}$ = 0.5, which is equal to the value at the solar neighbourhood (See \S 4) such that $\mathcal{T}_s = 1.15$. Therefore, we can say $\mathcal{T}_i Q_i$ is the thickness-corrected stability parameter, $Q_i$ being the stability parameter for the $i^{th}$ component as given in Equation (1). The thickness corrections are expected to be robust and not dependent on the choice of the 
vertical density profile or the oblateness (Romeo 1992).

\section{Sample} We choose 5 LSBs with inclination $i < 40^o$ from the sample of de Blok et al. (2001), who obtained mass models for a sample of LSBs, using high
 resolution HI 21cm radio-synthesis observations and R-band photometry. Our choice was guided by the fact that stellar photometry of nearly face-on galaxies are less 
affected by dust obscuration than those of their more edge-on counterparts, and, therefore, closely represents the true underlying stellar population even 
without extinction correction. Our LSB sample consist of F563-1, F568-1, F568-3, F568-VI and F579-VI. They have a B-band central surface surface brightness ${{\mu}_B}(0)$ ranging between 23-24 mg arcsec$^{-2}$, spanning exponential disc scalelength $R_D$ between 2.8 - 5.3 kpc and comparable dynamical masses as given by the asymptotic
rotational velocity V$_{\rm{max}}$ $\sim$ 105 - 118 kms$^{-1}$, except for F568-1 which has a slightly higher V$_{\rm{max}}$ $\sim$ 142 kms$^{-1}$. 
All our galaxies are bulgeless. 
 Our sample LSBs have a central HI surface density $\sim$ a few M$_{\odot}$pc$^{-2}$. The HI surface density profile of 
only F568-1 clearly indicates the presence of a hole at its centre; for the rest of the galaxies, the HI surface density remains almost constant before falling off with 
radius. In Table 1, we summarize the basic properties of our sample galaxies. 

\begin{table}
	\centering
	\begin{minipage}{150mm}
	\caption{Basic properties of our sample of LSBs}
	\label{tab:example_table}
	\begin{tabular}{lccccc} 
		\hline
		Name\footnote{All values quoted from de Blok et al. (2001)} & D\footnote{Distance} & ${{\mu}_B}(0)$\footnote{B-band central surface brightness} & $R_D$\footnote{Exponential disc scalelength} & i\footnote{Inclination} & $V_{\rm{max}}$\footnote{Maximum Velocity}\\
                     &(Mpc) & (mag arcsec$^{-2}$) & (kpc)& ($^o$) & (kms$^{-1}$) \\
		\hline
		F563-1 &45 & 23.6 &2.8 & 25 & 112 \\
		F568-1 &85 & 23.8 &5.3 & 26&  142\\
		F568-3 &77 & 23.1 &4.0 &40& 105\\
		F568-VI &80 & 23.3  &3.2&40 &118  \\
		F579-VI &85 & 22.8 &5.1 &26 &114  \\
		
		\hline
	\end{tabular}
\end{minipage}
\end{table}

\section{Input Parameters}

According to Equation (1) - (4), the input parameters required to calculate Q$_{\rm{RW}}$ are 
the epicyclic frequency ($\kappa$), the stellar and the gas surface densities (${\mu}_{s}$, ${\mu}_{g}$) and the 
stellar and the gas velocity dispersions (${\sigma}_{R,s}$, ${\sigma}_{R,g}$). 

The epicyclic frequencies were estimated as follows: a Brandt profile (Brandt 1960)
was fitted to the observed rotation curve as given in de Blok et al. (2001), which
gave ${\Omega}_{\rm{Total}}$ as a function of R in an analytical form. $\kappa$ could then be estimated using Equation (3).

The stellar surface densities ${\mu}_{s}$ were determined from the R-band central surface brightness values, 
the exponential disc scalelengths and a constant mass-to-light ratio of 1.4 as given in de Blok et al. (2001).
The color variation with galacto-centric radius is found to be modest in LSB galaxies, and hence the assumption of 
a mass-to-light ratio constant with radius is justified.

The gas surface densities ${\mu}_{g}$ 
were obtained by multiplying the HI surface densities given in de Blok, McGaugh and van der Hulst (1996) by 1.4 to take into
 account corrections due to the presence of He and other metals. 
de Blok et al. (2001) did not taken into account
the presence of molecular gas $H_2$ in their mass models. This is a reasonable assumption
as $H_2$ may not be a dynamically significant component in LSBs, as the $H_2$ to HI
mass ratio in late-type LSB spirals was found to be low $\sim$ 10$^{-3}$ (Schombert et al. 1990, Matthews et al. 2005). Therefore, the contribution
of $H_2$ can be safely ignored in mass models of LSBs, without incurring significant errors.

No direct observational estimates were available for the radial stellar velocity dispersion ${\sigma}_{R,s}$ for most of our sample LSBs. 
Therefore ${\sigma}_{R,s}$ were determined by first modelling the vertical velocity dispersion using 
the 2-component galactic disc model of gravitationally-coupled stars and gas in the force field of the dark matter halo, and taken to be in vertical hydrostatic equilibrium (See, for example, Banerjee \& Jog 2007). It was further assumed that the stellar disc had an average scaleheight of $R_D/6$ ($R_D/5$ - $R_D/7$) and that the dispersion declined exponentially with $R$ with a scalelength equal to 2 $R_D$ (van der Kruit \& Searle 1981); the radial velocity dispersion was then determined assuming a vertical-to-planar stellar velocity dispersion ratio of $\sim$ 0.5, which is the value at the solar neighbourhood (See, for example, Binney \& Tremaine 1987). 
However, Gerssen \& Shapiro Griffin (2012) showed that vertical-to-planar stellar velocity dispersion ratio decreases markedly from
early- to late-type galaxies. For an Sbc galaxy like the Milky Way, this value is $\sim$ 0.5.
 But for later-type galaxies like the LSBs, it can be significantly smaller, closer to 0.3.
Assuming a value of 0.3 for the vertical-to-planar stellar velocity dispersion ratio 
would result, if at all, in a slightly higher value for Q$_{\rm{RW}}$, which, as will be seen, will further strengthen the main conclusions of the paper,
and, therefore, our choice of a vertical-to-planar stellar velocity dispersion ratio of ~ 0.5 is rather conservative. 
 However, as will also be seen later in the section, the dependence of Q$_{\rm{RW}}$  on the ${\sigma}_{R,s}$ term is rather weak. In fact, we checked that a change in the value of ${\sigma}_{R,s}$ by upto $\sim$ 30 \% 
hardly changes the calculated value of Q$_{\rm{RW}}$. 
In Figure 1 [Left Panel], we compare our model-derived values of the central ${\sigma}_{R,s}$ for our sample LSBs, with those of the LSBs in the sample of 
Martinsson et al. (2013) who directly obtained the same from integral field spectroscopy studies. We note that within error bars, our model-predicted values for the central ${\sigma}_{R,s}$ are mostly consistent with the directly measured values in other LSBs .
In Figure 1 [Right Panel], we present the ratio of the scalelength for the exponential fall-off of the velocity dispersion in the radial direction $h_{\sigma}$ to the 
exponential disc scalelength $R_D$ as obtained for the LSB sample in Martinsson et al. (2013). We find that $h_{\sigma}/R_D$ $\sim$ 1 - 3 for most of the galaxies,
 instead of being 2 as assumed by us in the current work. However, two of our sample galaxies, F563-1 and F568-3, were 
studied by Kuzio de Naray, McGaugh and de Blok (2008), who obtained high-resolution optical velocity fields using integral field spectroscopy for a sample of 
LSBs, and estimated the non-circular motion to be $\sim$ 25-30 kms$^{-1}$. Interestingly, we found that the Q$_{\rm{RW}}$ value remains almost unchanged 
if we use a constant ${\sigma}_{R,s}$ $\sim$ 30 kms$^{-1}$ instead of the model-predicted values 
, again implying a weak dependence 
of Q$_{\rm{RW}}$ on the  ${\sigma}_{R,s}$ value. 
This is possibly because in a gas-rich system like an LSB, the disc dynamics is equally regulated by the stellar and gas 
component at almost all radii (Banerjee, Matthews \& Jog 2010), unlike in normal spirals where the stellar gravity plays the key role in regulating disc dynamics.
As a result, the calculated value of Q$_{\rm{RW}}$ is effectively not so sensitive to the moderate change in value of ${\sigma}_{R,s}$.

\begin{figure*}
\begin{center}
\begin{tabular}{cccc}
\resizebox{75mm}{!}{\includegraphics{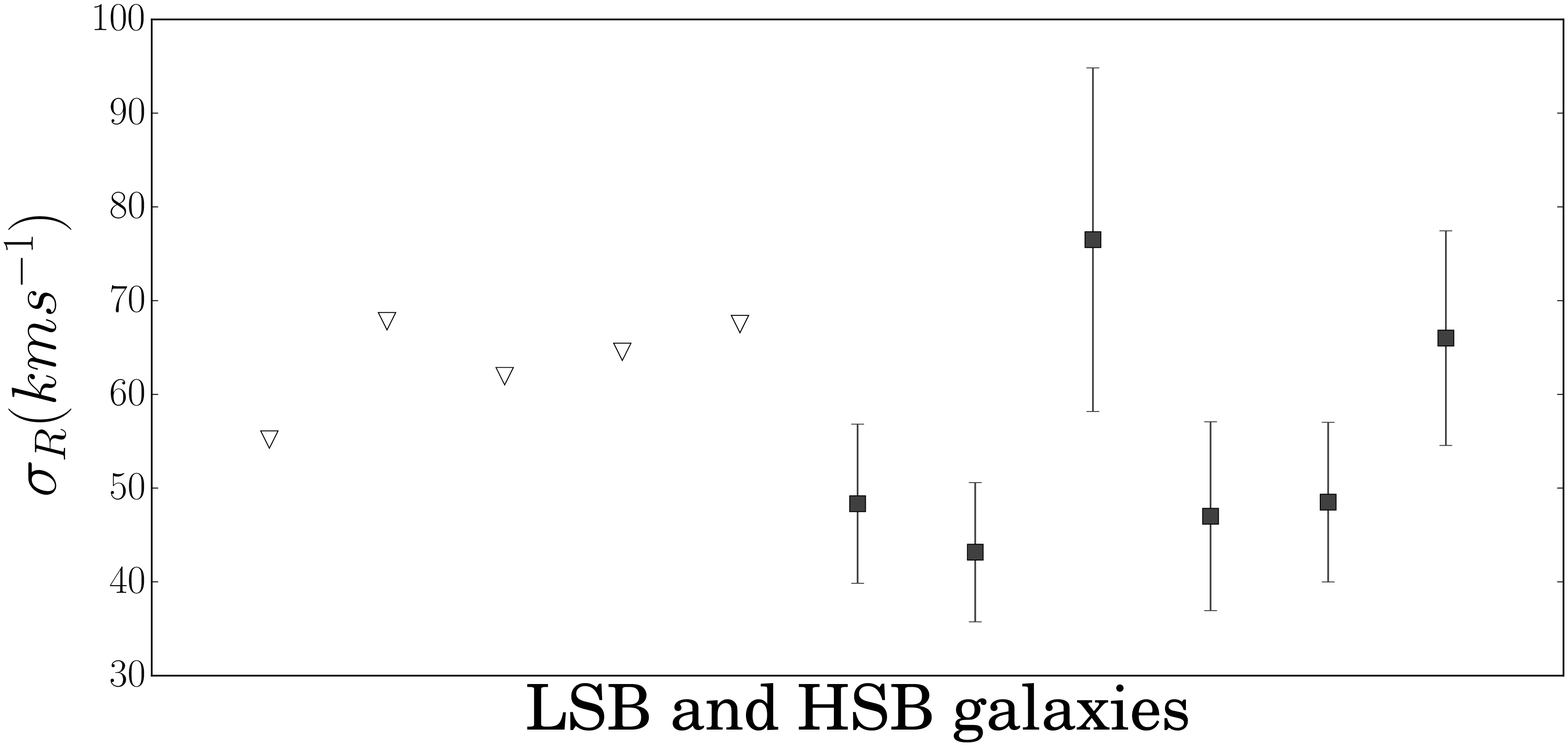}} &
\resizebox{75mm}{!}{\includegraphics{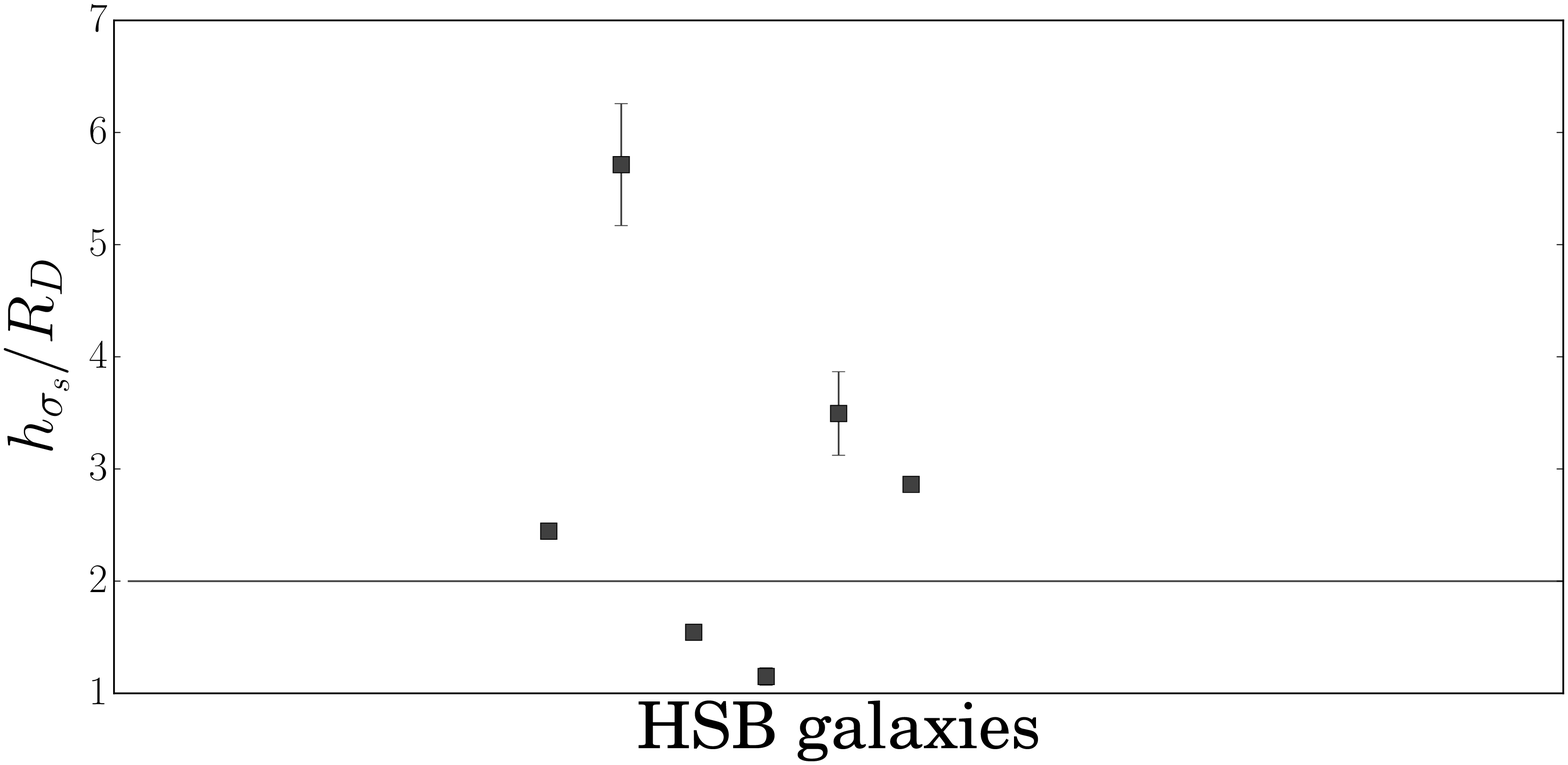}} \\
\end{tabular}
\end{center}
\caption{[Left Panel] Comparison of ${\sigma}_{R,s}(0)$ , the central stellar velocity dispersion for our sample LSBs 
 as estimated using our dynamical model (open inverted triangles), with those of the LSBs in the sample of Martinsson et al. (2013), 
as determined from integral field spectroscopy studies (filled squares). [Right Panel] $h_{\sigma}/R_D$ , the ratio of the scalelength for 
the exponential fall-off in the stellar velocity dispersion in the radial direction to the exponential disc scalelength, as obtained for a sample of LSBs from integral 
field spectroscopy studies by Martinsson et al. (2013). The horizontal line represents $h_{\sigma}/R_D$ = 2, as assumed in our dynamical model to estimate the 
stellar velocity dispersion for our sample LSBs. }
\label{fig:nfw_5249}
\end{figure*}

Similarly, our sample LSBs did not have directly measured values for the gas velocity dispersion ${\sigma}_{R,g}$.
However, recent studies of two edge-on LSBs (UGC7321,IC5249) among a sample of edge-ons (O'Brien, van der Kruit \& Freeman 2010c)
 revealed that ${\sigma}_{R,g}$ varies between 7 - 10 kms$^{-1}$ over most of the disc, which is also roughly consistent
 with the measurements in other galaxies in general; 6.5 $\pm$ 1.5 kms$^{-1}$ (narrow component of HI) and 16.8 $\pm$ 4.3 (cold component of HI) 
(Ianjamasimanana et al. 2012), and 6.5 - 10.5 kms$^{-1}$ (Westfall et al. 2014).
In this paper, we consider two different cases for all our analyses: ${\sigma}_{R,g}$ remaining constant with radius at 7 kms$^{-1}$ and 10 kms$^{-1}$ respectively.
This should be a conservative choice, given our aim of comparing the disc stabilities of LSBs and HSBs.

\section{Results \& Discussion}


We first calculate the theoretical disc stabilities of our sample galaxies as a function of galactocentric radius R.
Figure 2 shows the 2-component disc stability parameter Q$_{\rm{RW}}$ versus $R$ for each of our sample  
LSB galaxies calculated using Equation (1). The open and the filled circles represent 
Q$_{\rm{RW}}$ obtained using ${\sigma}_{R,g}$ =  7 kms$^{-1}$ and 10 kms$^{-1}$ respectively, which should span the whole range of ${\sigma}_{R,g}$ values
as constrained by direct observational studies of LSBs and other spirals (See \S 4). The error bars on our calculated Q$_{\rm{RW}}$ values have been estimated by propagating the error bars on the different input parameters, the dominant contributions being from the $\kappa$ and the ${\mu}_g$ terms, thus reflecting the
 error bars on the rotation curve and the HI surface density plots. In fact, as will be discussed later in this section, Q$_{\rm{RW}}$ has a moderately strong dependence on ${\sigma}_{R,g}$, and hence any error in the value of ${\sigma}_{R,g}$ should have been strongly reflected in the error in Q$_{\rm{RW}}$. However, in this paper, we have considered ${\sigma}_{R,g}$ as an input parameter, and, therefore, held it constant at a fixed value, 7 kms$^{-1}$ or 10 kms$^{-1}$, which are, respectively, the minimum and maximum possible value of ${\sigma}_{R,g}$ as observed in LSBs. Therefore, for each of our sample LSBs, the Q$_{\rm{RW}}$ plots corresponding to ${\sigma}_{R,g}$ =  7 kms$^{-1}$ and 10 kms$^{-1}$ respectively, roughly determine the range of Q$_{\rm{RW}}$ values at a given R.
The dotted horizontal line corresponds to
Q$_{\rm{RW}}$ = 1, which represents a disc marginally stable against local, axi-symmetric perturbations; Q$_{\rm{RW}}$ > 1 indicates an stable disc and vice-versa.
We note that for all our sample LSBs in general, Q$_{\rm{RW}}$ first falls off, then remains almost constant till R $<$ 3 R$_D$, and 
finally rises with R, for R $>$ 3 R$_D$. Interestingly, 3 R$_D$ also roughly marks the periphery of the stellar disc of a galaxy.
The location of the minima in Q$_{\rm{RW}}$ as well as its width vary from galaxy to galaxy.
This is similar to the general trend found in the combined sample of HSBs and LSBs in Westfall et al. (2014) [Figure 2 of Westfall et al. 2014]. 
As dicussed in \S 2, the disc stability parameter Q$_{\rm{RW}}$ and therefore the shape of its radial profile, within the purview of the adopted dynamical model, is regulated by the radial profiles of the $\kappa$, ${\mu}_s$, ${\mu}_g$
and the ${\sigma}_s$ (We do not include ${\sigma}_g$ as it has been taken to remain constant with radius). 
It is easy to check that the shape of the Q$_{\rm{RW}}$ profile is determined by an interplay between the exponentially falling 
${\mu}_s$ + ${\mu}_g$,
and $\kappa$ falling off almost as 1/R; the dependence of Q$_{\rm{RW}}$ on the ${\sigma}_s$ has been found to be weak.
This radial trend in Q$_{\rm{RW}}$ is also roughly consistent with the radial distribution of the H-$\alpha$ emission line surface brightness, which is an indicator of
 the recent star-formation rate surface density, in a sample of disc galaxies studied by Kennicutt (1989) [Figure 2 of Kennicutt 1989], which 
already implies the crucial role played by the disc dynamical stability in regulating star formation in galaxy discs. 
Besides, we also note the strong dependence of Q$_{\rm{RW}}$ on the assumed value of ${\sigma}_{R,g}$.  
Besides, we find that as ${\sigma}_{R,g}$ changes from 7 to 10 kms$^{-1}$, Q$_{\rm{RW}}$ changes almost by the same factor at all R for all our sample galaxies, except for F579-VI. This anomaly in F579-VI can be explained by the fact that F579-VI has a dark matter halo, denser and more compact 
by almost an order of magnitude than those of other LSBs in the sample, with similar asymptotic rotational velocities 
[Table 6 of de Blok et al. 2001].
As a result, the disc dynamics in this galaxy is effectively regulated by the underlying gravitational potential of the dark matter halo, as is also evident from its rotation curve decomposition into the disc and the dark matter halo components [Figure 3 of de Blok et al. 2001]; thus the role of 
the other input parameters, including gas velocity dispersion, is rendered insignificant in regulating the Q$_{\rm{RW}}$ values of F579-VI. 
In Table 2, we summarize Q$_{\rm{RW}}^{\rm{min}}$, which is the minimum Q$_{\rm{RW}}$ in 0.1 $\le$ R/R$_D$ $\le$ 2.5, 
for our sample LSBs; Q$_{\rm{RW}}^{\rm{min}}$ may also be taken to be a conservative estimate of the Q$_{\rm{RW}}$ for the galactic disc,
 as proposed by Westfall et al. (2014).
We note that for relatively cold galactic discs (${\sigma}_{R,g}$ =  7 kms$^{-1}$), the Q$_{\rm{RW}}$ value may vary significantly 
($\sim$ 32 $\%$) from galaxy to galaxy; for hotter discs (${\sigma}_{R,g}$ =  10 kms$^{-1}$), the variation in Q$_{\rm{RW}}$ is much less ($\sim$ 17 $\%$),
reflecting that the effect of the other physical parameters govering disc stability is relatively suppressed in a hotter galactic disc.
We choose the sample median as representing the average value of the Q$_{\rm{RW}}^{\rm{min}}$ for our sample LSBs. 
We find a median  Q$_{\rm{RW}}^{\rm{min}}$ = 2.6 with ${\sigma}_{R,g}$ = 7 kms$^{-1}$ , and 
Q$_{\rm{RW}}$ = 3.1 with ${\sigma}_{R,g}$ = 10 kms$^{-1}$. 
In comparison, the Toomre Q value corresponding to the observed HI surface density, which sets the star 
formation threshold in a large sample of HSBs, is 1.4 (Kennicutt 1989). This disc stability value was however modelled by approximating the disc  
as a single component system with zero vertical thickness, which may not be a good assumption in general; the corresponding value for the 2-component disc stability, which better represents the disc stability for a star-gas system, may be even lower (See \S 1 for details). This already indicates that LSBs have higher disc stability than normal HSBs. Interestingly, our calculations show that the disc instability is not driven by any particular disc component (i.e., stars or gas) for
all our sample galaxies. Even for a given LSB galaxy, the disc instability may be driven by the stellar component at some $R$, the gas component at others, and
 equally by both the components at certain $R$; also, the trend may vary from galaxy to galaxy. This is unlike the case of normal galaxies where stars are found to be the the primary driver of gravitational instabilities (Romeo \& Magotsi 2017). Besides, it was also shown that this feature holds good even in the inner disc of NGC 1068, a powerful nearby Seyfert+starburst galaxy (Romeo \& Fathi 2016).This is possibly the reflection of the fact that in gas-rich LSB galaxies, the self-gravities of both the stellar and the gas component dominate the disc dynamics on almost an equal footing, unlike in normal spirals where the stellar component mainly contributes to the net gravitational potential.

\begin{table}
	\centering
	\caption{Q$_{\rm{RW}}^{\rm{min}}$ of our sample of LSBs}
	\label{tab:example_table}
	\begin{tabular}{lcc} 
		\hline
		Name &  Q$_{\rm{RW}}^{\rm{min}}$ & Q$_{\rm{RW}}^{\rm{min}}$\\
		     &  (${\sigma}_{R,g}$ =  7 kms$^{-1}$) & (${\sigma}_{R,g}$ =  10 kms$^{-1}$) \\
        	\hline
		F563-1 & 2.8 $\pm$ 0.2 & 3.4 $\pm$ 0.3\\
		F568-1 & 2.2 $\pm$ 0.1 & 2.9 $\pm$ 0.1\\
		F568-3 & 2.0 $\pm$ 0.1 & 2.3 $\pm$ 0.1\\
		F568-VI & 2.6  & 3.2\\
		F579-VI & 2.9  & 3.1\\
		
		\hline
	\end{tabular}
\end{table}

We now compare the Q$_{\rm{RW}}^{\rm{min}}$ values for our sample LSBs with those of the combined sample of
LSBs and HSBs as was obtained by Westfall et al. (2014). However, the B-band central surface brightness were not available in the literature for all the
galaxies in the above sample. 
Hence, we could not identify LSBs and HSBs in their sample, 
  based on the definition of LSBs involving their B-band central surface brightness values (See \S 1) as is the usual norm, and has been done so far in this paper.
The R-band central surface brightness values of these galaxies were, however, given in Bershady et al. (2010). 
We have therefore resorted to an alternative definition of LSBs, based on their R-band central surface brightness, ${\mu}_R$(0), to distinguish between LSBs and
 HSBs in their sample. According to this definition, LSBs are galaxies with ${\mu}_R$(0) < 20.8 mag arcsec$^{-2}$ and vice versa.
(Courteau et al. 1996; Adami et al. 1996). 
Based on this definition, we find that the sample LSBs of Westfall et al. (2014) have a median Q$_{\rm{RW}}^{\rm{min}}$ = 2.2 $\pm$ 0.4,
whereas the HSBs have a median Q$_{\rm{RW}}^{\rm{min}}$ = 1.8 $\pm$ 0.3.
This means, that although the median Q$_{\rm{RW}}^{\rm{min}}$ value for the LSBs is slightly higher than that of the HSBs,  
within error bars, the disc stability values of the LSBs and HSBs are comparable according to Westfall et al. (2014). 
In comparison, the median Q$_{\rm{RW}}^{\rm{min}}$ value for our LSB sample (even for the ${\sigma}_{R,g}$ =  7 kms$^{-1}$ case) is clearly higher:
2.6, than their of the HSB sample. 
This may be understood by the fact that the stellar surface density values used by Westfall et al. (2014)
were possibly over-estimated. 
The dynamical surface density was modelled approximating the galaxy as a single component system of self-gravitating stars, thereby
 ignoring the self-gravity of the gas and the dark matter halo. However, the contribution of these components may not be negligible. In fact,
 although gas constitutes only a sub-dominant mass fraction of the galactic disc, it nevertheless strongly regulates the disc dynamics, being confined to a thin layer near the galactic midplane, as it has a relatively lower value of the vertical velocity dispersion compared to the stars (Banerjee \& Jog 2007);
 this effect will be especially significant in gas-rich, late-type systems like the LSBs.
Besides, dark matter has been found to dominate the disc gravitational potential at all radii in LSBs (Banerjee et al. 2010), and therefore the presence of dark matter
can not be ignored even at small and intermediate radii unlike in HSBs. 
However, in the model adopted by Westfall et al. (2014), the stellar surface density was taken to be equal to the dynamical surface density less the gas surface density value, thus
 neglecting the contribution of the dark matter halo twice. Therefore, the stellar surface density values of Westfall et al. (2014) appear to have been
over-estimated, which possibly resulted in under-estimating disc stability values i.e., lower values of Q$_{\rm{RW}}^{\rm{min}}$, especially for the 
LSBs. Having said that, we may conclude, Q$_{\rm{RW}}^{\rm{min}}$ value for LSBs are on an average
higher than those of HSBs, thus confirming that the discs of LSBs are dynamically more
stable against star formation compared to those of HSBs (Also see, Ghosh \& Jog 2014).
In Figure 3, we compare the Q$_{\rm{RW}}^{\rm{min}}$ values of the LSBs and HSBs in the sample of Westfall et al. (2014) with those calculated in
 this paper for our sample LSBs. Before concluding this discussion, we may note that Q$_{\rm{RW}}$ = 1 gives the critical stability level of a galactic disc in the presence of local, axi-symmetric perturbations only. However, realistic galactic discs are by no means subjected to local, axisymmetric perturbations only.
 In fact, non-axisymmetric perturbations lead to the formation of bars and spiral arms in galactic discs. Griv \& Gedalin (2012) showed that non-axisymmetric perturbations have a destabilizing effect and therefore may increase the critical stability level for local axisymmetric perturbations to Q$_{\rm{RW}}$ $\sim$ 1 - 2(Romeo \& Fathi 2015). In addition to this, there may be gas dissipation effects, which may raise the critical stability level further to Q$_{\rm{RW}}$ $\sim$ 2-3 (Elmegreen 2011). The Q$_{\rm{RW}}^{\rm{min}}$ values for all our sample LSBs clearly lies between 2-3, which confirms that LSB discs are stable against local perturbations under realistic conditions of the inter stellar medium (ISM), when non-axisymmetric perturbations and gas dissipation are also at play. In contrast, the 
median Q$_{\rm{RW}}^{\rm{min}}$ for HSB discs, as mentioned above, is $\sim$ 1.8 indicating HSB discs are not quite stable against local, non-axisymmetric perturbations under realistic conditions, unlike their LSB counterparts.
 
\begin{figure*}
\begin{center}
\begin{tabular}{cccc}
\resizebox{150mm}{!}{\includegraphics{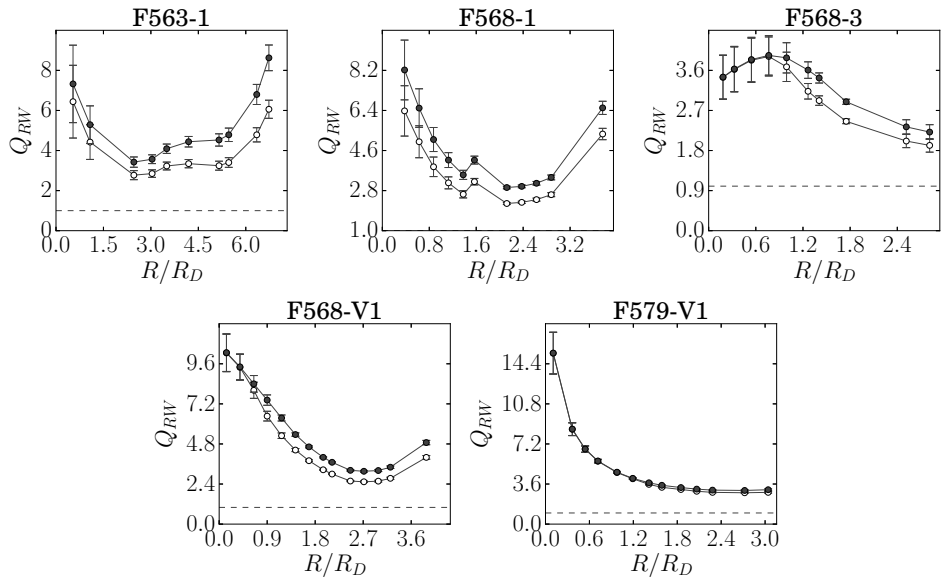}} \\
\end{tabular}
\end{center}
\caption{2-component disc stability parameter Q$_{\rm{RW}}$, proposed by Romeo \& Wiegert (2011), as a function of galactocentric radius $R$, as calculated 
in this paper for a sample of 5 LSBs. The open and the filled circles represent
Q$_{\rm{RW}}$ calculated using ${\sigma}_{R,g}$ =  7 kms$^{-1}$ and 10 kms$^{-1}$ respectively. The dotted horizontal line indicates
Q$_{\rm{RW}}$ = 1, which represents
a marginally stable disc}
\label{fig:nfw_5249}
\end{figure*}

We next check if the calculated Q$_{\rm{RW}}$ values for our sample LSBs, which is expected to capture the essential physics driving star formation according to the
 purely dynamical model adopted here, actually correlate with their star formation rate, as directly determined from observational studies. There are several tracers of star formation rates in galaxies (See, for example, Kennicutt \& Evans 2012 for a review); UV surface brightness, for instance, is a probe of the recent star formation activity. Wyder et al. (2009) determined the global star formation rates of a sample of LSB galaxies using their UV surface brightness, and their sample contained all five of our sample LSBs. In Figure 4, we study the star formation rate surface density [adapted from Wyder et al. 2009] as a function of Q$_{\rm{RW}}^{\rm{min}}$, as calculated in this paper, for our sample LSBs. We find that the data points can be
 well-fitted by a power-law with a slope of -1.50 $\pm$ 0.59, thus confirming that disc dynamical stability (Q$_{\rm{RW}}^{\rm{min}}$) primarily regulates star formation rate density in galaxies. Also, our best-fitting slope roughly matches a slope of $\sim$ -1.5 found in hydrodynamical simulations studying
 the dependence of star formation rate on disc stability of ordinary spirals in general (Li et al. 2006). This further implies that the disc dynamical stability 
 regulates the star formation rate densities in galaxies equally strongly, irrespective of galaxy type or morphology;
the lower star formation rates exhibited by LSBs is simply the outcome of their higher disc dynamical stabilities. However, Westfall et al. (2014), reported a steeper slope of $\sim$ -3 for their combined sample of HSBs and LSBs, which is puzzling. But it is possibly related to the over-estimated disc surface density
 values used in determining Q$_{RW}^{\rm{min}}$, as discussed earlier.
Besides, they used a different diagnostic for recent star formation viz. integrated 21 cm continuum luminosity, to estimate their global star formation rates, 
which may also explain the observed difference in the power-law slopes found in the two studies. The causal connection between disc instability and star formation was first suggested by Kennicutt (1989), which was founded on the basic fact that disc instabilities trigger formation of gas clumps, the coalescence of which eventually leads to star formation. However, the above connection is not direct, and may be mediated by several factors, the strongest among them being feedback. Besides, it has been recently found that formation of gas clumps may also result in the heating of ISM in addition to the gas depletion in their neighbourhood, which, in turn, are likely to oppose star formation to some extent, and thus also act as a feedback effect. Therefore, we stress the fact that inspite of our above finding, the connection between growth of disc instabilities and star formation  is not quite direct and can be better understood by factoring in feedback effects in general (Kennicutt \& Evans 2012, Forbes et al. 2014, Goldbaum et al. 2015, 2016). Such a study is, however, outside the aim and scope of the current paper.

\begin{figure}
        \includegraphics[width=\columnwidth]{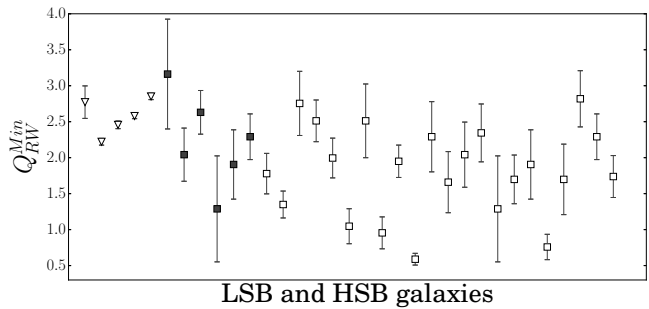}
    \caption{Comparison of Q$_{\rm{RW}}^{\rm{min}}$, the minimum value of the 2-component disc stability parameter proposed by 
Romeo \& Wiegert (2011), as calculated in this paper for a sample of 5 LSBs (open inverted triangles), with those for a sample of HSBs (filled squares) and 
LSBs (open squares) as given in Westfall et al. (2014).}
    \label{fig:Qsg_min_dm}
\end{figure}

\begin{figure}
        \includegraphics[width=\columnwidth]{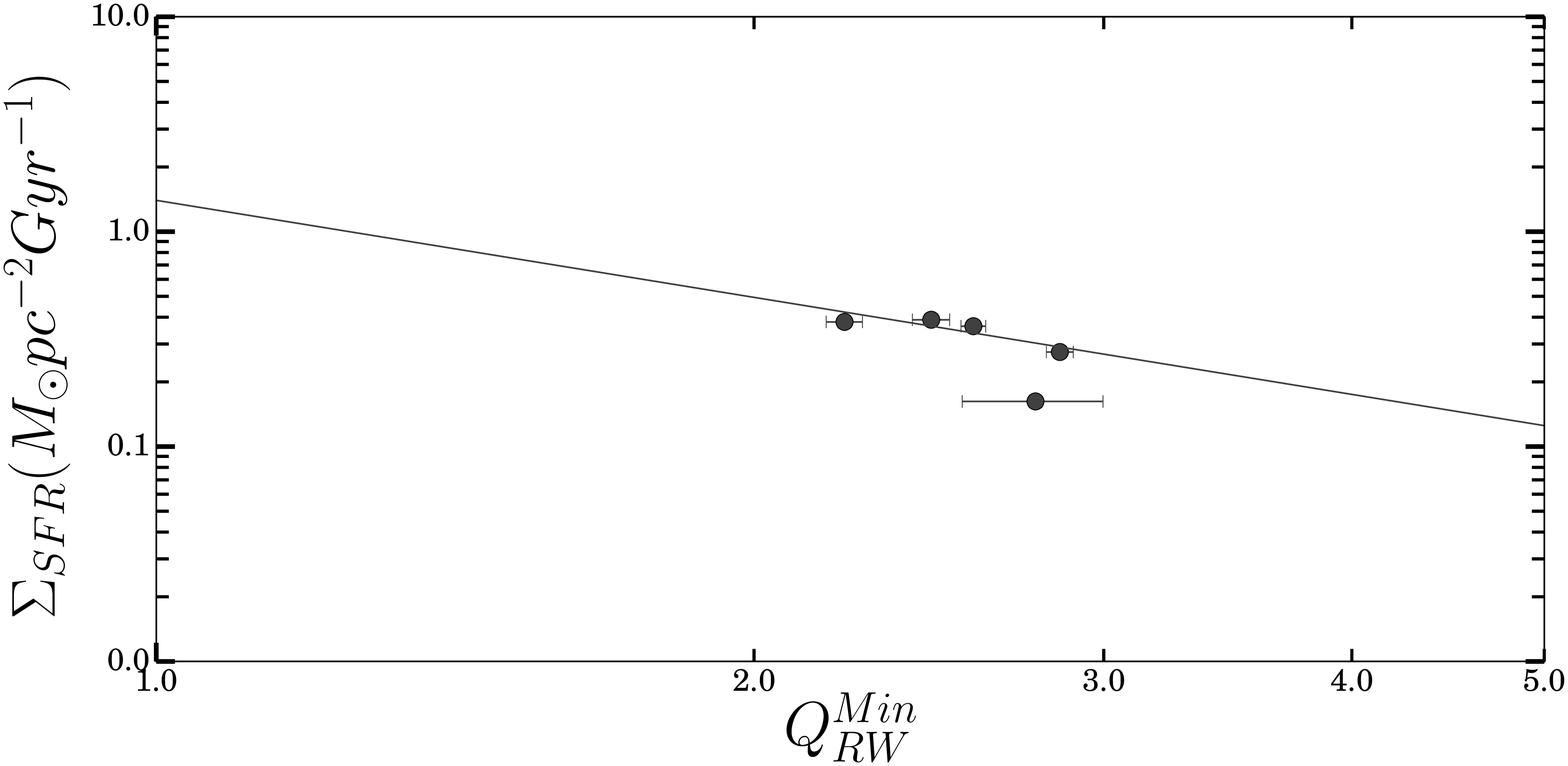}
    \caption{${\Sigma}_{\rm{SFR}}$, the star formation rate surface density (adapted from Wyder et al. 2009) as a function of 
Q$_{\rm{RW}}^{\rm{min}}$, 
the minimum value of the 2-component disc stability parameter proposed by Romeo \& Wiegert (2011), as calculated in this paper for a sample of 5 LSBs. 
The straight line shows the best-fitting power-law dependence of ${\Sigma}_{\rm{SFR}}$ on Q$_{\rm{RW}}^{\rm{min}}$, the power-law slope being -1.50 $\pm$ 0.59.}
    \label{fig:Qsg_min_dm}
\end{figure}

\begin{figure*}
\begin{center}
\begin{tabular}{cccc}
\resizebox{150mm}{!}{\includegraphics{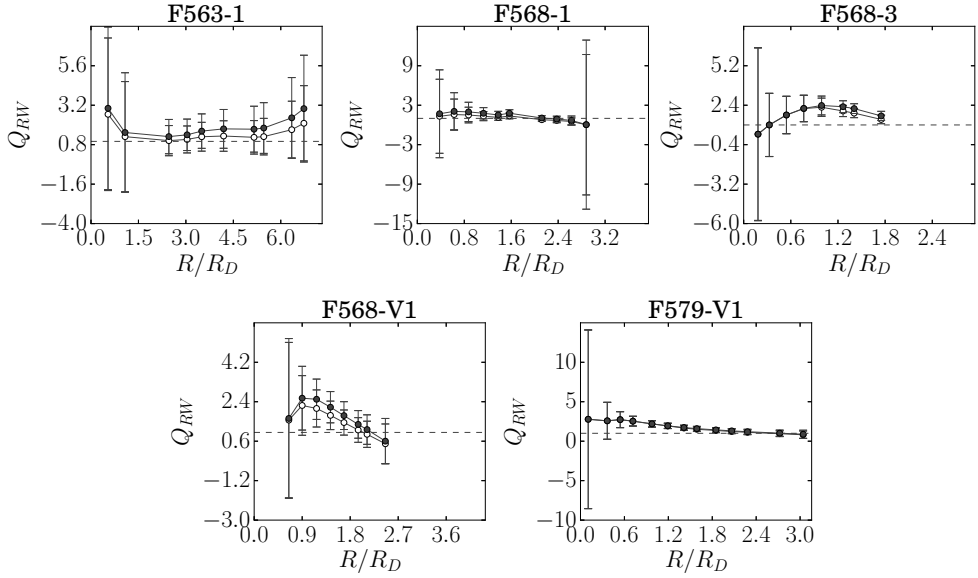}} \\
\end{tabular}
\end{center}
\caption{2-component disc stability parameter Q$_{\rm{RW}}$, as proposed by Romeo \& Wiegert 2011, as a function of galactocentric radius $R$, as calculated in this paper for a sample of 5 LSBs, but in the presence of the 
gravitational potential due to the disc (stars + gas) alone  i.e., ignoring the gravitational potential due to
 the dark matter halo . The open and the filled circles represent
Q$_{\rm{RW}}$ calculated using ${\sigma}_{R,g}$ =  7 kms$^{-1}$ and 10 kms$^{-1}$ respectively. The dotted horizontal line indicates
Q$_{\rm{RW}}$ = 1, which represents a marginally stable disc.}
\label{fig:nfw_5249}
\end{figure*}

\begin{figure*}
\begin{center}
\begin{tabular}{cccc}
\resizebox{150mm}{!}{\includegraphics{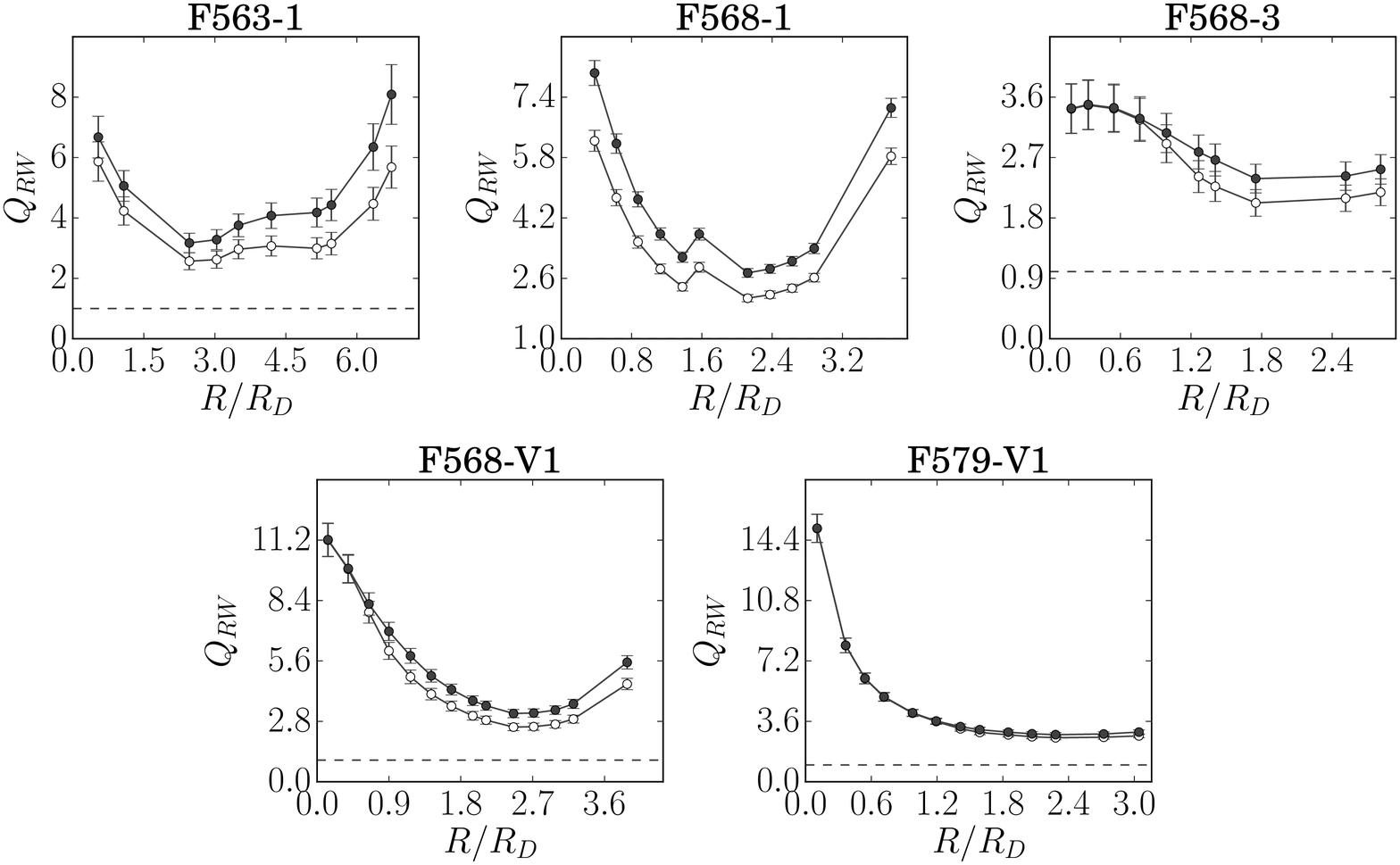}} \\
\end{tabular}
\end{center}
\caption{2-component disc stability parameter Q$_{\rm{RW}}$, proposed by Romeo \& Wiegert 2011, as a function of galactocentric radius $R$, as calculated in this paper
for a sample of LSBs, but in the presence of the gravitational potential due to the dark matter halo alone i.e., ignoring the gravitational potential due to
 the self-gravity of the disc. The open and the filled circles represent Q$_{\rm{RW}}$ calculated using ${\sigma}_{R,g}$ =  7 kms$^{-1}$ and 10 kms$^{-1}$ respectively. The dotted horizontal line indicates Q$_{\rm{RW}}$ = 1, which represents a marginally stable disc. }
\label{fig:nfw_5249}
\end{figure*}

Finally, we aim to asses the relative roles of the disc and the dark matter halo in stabilizing the disc for our sample LSBs. 
The classic work of Ostriker \& Peebles (1973) demonstrated the necessity of an additional mass component, attributable either to the baryonic disc
 or to a dark matter halo, to stabilize a galactic disc having a velocity dispersion comparable to that of the Galaxy against global non-axisymmetric instabilities, which lead to bar formation.
As discussed in \S 4, ${\sigma}_{R,s}$ and ${\sigma}_{R,g}$ values for LSBs are almost comparable to those of HSBs. But since LSBs have relatively lower dynamical 
masses than HSBs, as indicated by their asymptotic rotational velocities, this velocity dispersion value may render the disc stable by itself.   
This apart, the disc may also be stabilized by the the centrifugal force from the coriolis spin-up of the perturbation, as contained in the epicyclic frequency term 
$\kappa$; in fact, in the current model, dark matter halo regulates the disc stability through this term only. Depending on the mass density profile of the disc and dark matter halo, 
$\kappa$ may be higher or lower, and accordingly the disc dynamical stability regulated. 
Therefore, we next study the role of the dark matter halo in stabilizing the LSB disc against local axi-symmetric instabilities, which trigger star formation.
In Figure 5, we present the Q${_{\rm{RW}}}$ as a function of R if the dark matter halo were absent i.e., the disc responded to its self-gravity alone.
i.e., the rotation curve and hence the epicyclic frequency were obtained using the gravitational potential due to the self-gravity of the stars and 
gas only (Equation 3). It clearly shows, for all our sample galaxies, the Q${_{\rm{RW}}}$ values tend to be even less than 1 
 at some radii , indicating that the disc is not stable by itself in the absence of the dark matter halo.
Similarly, in Figure 6, we present the Q${_{\rm{RW}}}$ the disc would have if the self-gravity of the disc were ignored i.e, the stars and
 gas were taken to respond to the gravitational potential due to the dark matter halo only. 
In contrast to the disc-alone case, here the Q${_{\rm{RW}}}$ values are all > 1, thus confirming that the 
gravitational potential of the dark matter halo is crucial to stabilize the disc. 
Interestingly, for each of our sample galaxies, 
the radial profiles of Q${_{\rm{RW}}}$
calculated using the dark matter potential alone (Figure 6) and those obtained using 
the total (disc + dark matter halo) potential (Figure 2)
 are almost indistinguishable, further underscoring the fact that disc stability in LSBs is primarily regulated by the dark matter halo.  
In fact, the Q$_{\rm{RW}}^{\rm{min}}$ calculated taking into account the gravitational potential of the dark matter only 
(median Q$_{\rm{RW}}^{\rm{min}}$ $\sim$ 2.3  - 3) is 
almost equal (within error bars) to that obtained using the total gravitational potential of the disc and the dark matter halo 
(median Q$_{\rm{RW}}^{\rm{min}}$ $\sim$ 2.6 - 3.1), but more than a factor of two higher than the disc-only case 
 (median Q$_{\rm{RW}}^{\rm{min}}$ $\sim$ 0.7  - 1.5). 
This conclusively shows that the dark matter is fundamentally responsible in regulating the high value of the disc dynamical stability in LSBs.
(But, see, Mihos et al. 1997).

To check for the fact that our conclusions are robust and not dependent on our choice of the 2-component disc stability parameter, we obtained the disc stability
using an alternative 2-fluid disc stability parameter as proposed by Jog (1996), with correction for the finite scaleheight of the disc; this turned out to be
different from the Q$_{\rm{RW}}$ values by at the most 14$\%$, thus confirming that our conclusions are not model-dependent.

\section{Conclusions} We investigate the origin of the low star formation rate (SFR) and hence the low surface brightness nature of some spiral galaxies, 
known as low surface brightness galaxies (LSBs). From a purely dynamical point of view, a galactic disc has a low star formation rate if it is 
relatively more stable against the growth of local axi-symmetric instabilities. This disc dynamical stability may be indicated by a high value (>1)
 of the 
2-component disc stability parameter, Q$_{\rm{RW}}$ for example, as proposed by Romeo \& Wiegert (2011).
 We find that the median value of Q$_{\rm{RW}}^{\rm{min}}$, the minimum of Q$_{\rm{RW}}$ over $R$, 
are 2.6 and 3.1 respectively for assumed HI velocity dispersion ${\sigma}_g$ = 7 and 10 kms$^{-1}$                      
respectively for our sample of LSBs, which is clearly higher than the median value of 1.8 $\pm$ 0.3 for Q$_{\rm{RW}}^{\rm{min}}$ for a sample
 of high surface brightness galaxies (HSBs) as obtained in earlier studies; this possibly explains the low SFR densities in LSBs. Besides, the 
observationally-determined SFR surface density values for our sample LSBs, as obtained by Wyder et al. (2009), have a power-law dependence on 
Q$_{\rm{RW}}^{\rm{min}}$, with the power-law   
slope of  -1.50 $\pm$ 0.59,  matching the slope of $\sim$ -1.5 found in hydrodynamical simulations of star formation in spirals in general; 
this indicates that the disc dynamical stability regulates the SFR densities in galaxies equally, irrespective of the galaxy type and morphology. 
Finally, we show that the Q$_{\rm{RW}}$ (>1) parameter is mainly determined by the underlying gravitational potential of the dark matter halo, 
thus confirming that the dark matter halo is primarily responsible for the disc stability and hence low star formation rates in low surface brightness galaxies.

\section*{Acknowledgements} 
The authors would like to thank Kanak Saha and Nissim Kanekar for useful suggestions and discussion.
Prerak Garg would like to acknowledge IUCAA VSP programme, and Arunima Banerjee DST-INSPIRE Faculty Fellowship (IFA14/PH-101) 
for supporting this research. Finally, the authors thank the referee, Alessandro Romeo, for his insightful and detailed comments, which have
 greatly improved the quality of the paper.

\section{References}
	Adami C. et al. 2006, ApJ, 728, L47 \\
	Banerjee A., Jog C.J., 2007, Astrophysical Journal, 662, 335 \\
	Banerjee A., Matthews L. D., Jog C. J., 2010, NewA, 15, 89 \\
	Bershady M. A. et al. 2010, ApJ, 716, 198 \\
	Binney, J. Tremaine, S. 1987, Galactic Dynamics. Princeton Univ. Press, Princeton, NJ \\
	Boissier S., Prantzos N. 1999MNRAS.307..857 \\
	Brandt J.C. 1960. Ap. J. 131, 293 \\
	Forbes, J. C., Krumholz, M. R., Burkert, A., Dekel, A., 2014, MNRAS 438, 1552 \\
	Courteau S. 1996, ApJS, 103, 363 \\
	de Blok W. J. G., McGaugh S. S., van der Hulst, J. M.  1996, MNRAS. 283, 18 \\
	de Blok W. J. G., McGaugh Stacy S., 2001, AJ, 122, 2396 \\
	Elmegreen, B. G., 2011, ApJ, 737, 10 \\
	Freeman K. C., 1970, ApJ, 160, 811 \\
	Gerssen, J., Shapiro Griffin, K., 2012, MNRAS, 423, 2726 \\
	Griv, E., Gedalin, M., 2012, MNRAS, 422, 600 \\
	Ghosh S., Jog, C. J. 2014, MNRAS, 439, 929 \\
	Goldbaum, N. J., Krumholz, M. R., Forbes, J. C., 2015, ApJ, 814, 131 \\
	Goldbaum, N. J., Krumholz, M. R., Forbes, J. C. 2016, ApJ, 827, 28 \\
	Hunter D. A., Elmegreen B. G., Baker A. L.,1998, ApJ, 493, 595 \\
	Ianjamasimanana R., de Blok W. J. G., Walter F., Heald G. H., 2012, AJ, 144, 96 \\
	Impey C., Bothun G., 1997, ARA\&A, 35, 267 \\
	Kennicutt R. C., Jr., 1989, ApJ, 344..685 \\
	Kennicutt, R. C., Evans, N. J. 2012, ARA\&A, 50, 531 \\
	Kuzio de Naray R., McGaugh S. S., de Blok W. J. G., 2008, ApJ, 676, 920 \\
	Li Y., Mac Low M-M., Klessen R. S., 2006, ApJ, 639, 879 \\
	Martinsson T. P. K., Verheijen M. A. W., Westfall K. B., Bershady M. A., Andersen, D. R., Swaters, R. A., 2013, A\&A, 557, 131 \\
	Matthews L. D., Gao Y., Uson J. M., Combes F., 2005, AJ, 129, 1849 \\
	McGaugh S. S., Bothun, G. D., 1994, AJ, 107, 530 \\
	Mihos, J. C., McGaugh, S. S., de Blok, W. J. G. 1997, ApJ, 477, 79 \\
	O'Brien J. C., Freeman, K. C., van der Kruit, P. C. 2010c, A\&A, 515, 62 \\
	Ostriker, J. P., Peebles, P. J. E., 1973, ApJ, 186, 467 \\
	Pfenniger, D., Combes, F., Martinet, L., 1994, A\&A, 285, 79 \\
	Romeo, A. B., 1992, MNRAS, 256, 307 \\
	Romeo, A. B., Falstad, N., 2013, MNRAS, 433, 1389 \\
	Romeo, A. B., Fathi, K., 2015, MNRAS, 451, 3107 \\
	Romeo, A. B., Fathi, K., 2016, MNRAS, 460, 2360 \\
	Romeo, A. B., Mogotsi, K. M.,  2017, MNRAS, 469, 286 \\
	Romeo, A. B., Wiegert, J., 2011, MNRAS, 416, 1191 \\
	Sackett, P. D., 1997, ApJ, 483, 103 \\
	Schombert J. M., Bothun G. D., Impey C. D.; Mundy, L. G., 1990, 100, 1523 \\
	Schombert J. M., McGaugh S. S., Eder, J. A., 2001, AJ, 121, 2420 \\
	Toomre, A.,1964, ApJ, 139, 1217 \\
	van der Hulst J. M., Skillman E. D., Smith T. R., Bothun G. D., McGaugh S. S., de Blok W. J. G., 1993, AJ, 106, 548 \\
	van der Kruit P. C., Searle L., 1981, A\&A, 95, 105 \\
        Westfall K. B., Andersen D. R., Bershady, M. A., Martinsson T. P. K., Swaters R. A., Verheijen M. A. W. 2014, ApJ, 785, 43 \\
	Wyder T. K. et al., 2009, ApJ, 696, 1834 \\

\end{document}